\documentclass{article}

\usepackage{arxiv}

\usepackage[utf8]{inputenc} % allow utf-8 input
\usepackage[T1]{fontenc}    % use 8-bit T1 fonts
\usepackage{hyperref}       % hyperlinks
\usepackage{url}            % simple URL typesetting
\usepackage{booktabs}       % professional-quality tables
\usepackage{amsfonts}       % blackboard math symbols
\usepackage{nicefrac}       % compact symbols for 1/2, etc.
\usepackage{microtype}      % microtypography
\usepackage{lipsum}		% Can be removed after putting your text content

\usepackage{graphicx}
\usepackage{natbib}
\usepackage{doi}
\usepackage{amssymb}
\usepackage{color}
\usepackage{soul}
\usepackage{enumitem, balance}
\usepackage{url}
\usepackage{rotating}
\usepackage{mdframed}
\usepackage{fontawesome}
\usepackage{adjustbox}

\title{Rethinking Sustainability Requirements: \\ Drivers, Barriers and Impacts \\ of Digitalisation from the Viewpoint of Experts}

\author{ \href{https://orcid.org/0000-0002-0636-5663}{\includegraphics[scale=0.06]{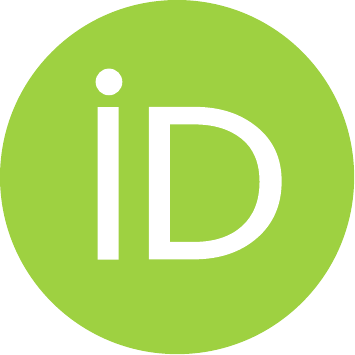}\hspace{1mm}Alessio Ferrari}
\\
	CNR-ISTI\\
	Pisa, Italy\\
	\texttt{alessio.ferrari@isti.cnr.it} \\
	\And
	\href{https://orcid.org/0000-0001-6733-1873}{\includegraphics[scale=0.06]{orcid.pdf}\hspace{1mm}Manlio Bacco} \\
	CNR-ISTI\\
	Pisa, Italy\\
	\texttt{manlio.bacco@isti.cnr.it} \\
	\And
	\href{https://orcid.org/0000-0002-6768-9047}{\includegraphics[scale=0.06]{orcid.pdf}\hspace{1mm}Kirsten Moore} \\
	Karlsruhe Institute of Technology\\
	Karlsruhe, Germany\\
	\texttt{kirsten.gaber@kit.edu} \\
	\And
	\href{https://orcid.org/0000-0003-3590-6331}{\includegraphics[scale=0.06]{orcid.pdf}\hspace{1mm}Andreas Jedlitschka} \\
	Fraunhofer IESE\\
	Kaiserslautern, Germany\\
	\texttt{andreas.jedlitschka@iese.fraunhofer.de} \\
	\And
	\href{https://orcid.org/0000-0003-4132-272X}{\includegraphics[scale=0.06]{orcid.pdf}\hspace{1mm}Steffen Hess} \\
	Fraunhofer IESE\\
	Kaiserslautern, Germany\\
	\texttt{steffen.hess@iese.fraunhofer.de} \\
	\And
	\href{https://orcid.org/0000-0001-6400-552X}{\includegraphics[scale=0.06]{orcid.pdf}\hspace{1mm}Jouni Kaipainen} \\
	University of Jyväskylä\\
    Kokkola, Finland\\
	\texttt{jouni.kaipainen@chydenius.fi} \\
	\And
	\href{https://orcid.org/0000-0001-5583-5396}{\includegraphics[scale=0.06]{orcid.pdf}\hspace{1mm}Panagiota Koltsida} \\
	ATHENA RC, \\National and Kapodistrian University of Athens\\
	Athens, Greece\\
	\texttt{p.koltsida@di.uoa.gr} \\
	\And
	\href{https://orcid.org/0000-0002-3907-2496}{\includegraphics[scale=0.06]{orcid.pdf}\hspace{1mm}Eleni Toli} \\
	ATHENA RC, \\National and Kapodistrian University of Athens\\
	Athens, Greece\\
	\texttt{elto@di.uoa.gr} \\
	\And
	\href{https://orcid.org/0000-0003-2905-9738}{\includegraphics[scale=0.06]{orcid.pdf}\hspace{1mm}Gianluca Brunori} \\
	Università degli Studi di Pisa, DISAAA\\
	Pisa, Italy\\
	\texttt{gianluca.brunori@unipi.it} 	
}

\renewcommand{\shorttitle}{\textit{arXiv} Template}

\hypersetup{
pdftitle={A template for the arxiv style},
pdfsubject={q-bio.NC, q-bio.QM},
pdfauthor={David S.~Hippocampus, Elias D.~Striatum},
pdfkeywords={First keyword, Second keyword, More},
}

\begin{document}
\maketitle

\begin{abstract}
Requirements engineering (RE) is a key area to address sustainability concerns in system development. Approaches have been proposed to elicit sustainability requirements from interested stakeholders before system design. However, existing strategies lack the proper high-level view to deal with the societal and long-term impacts of the transformation entailed by the introduction of a new technological solution. This paper proposes to go beyond the concept of system requirements and stakeholders' goals, and raise the degree of abstraction by focusing on the notions of \textit{drivers}, \textit{barriers} and \textit{impacts} that a system can have on the environment in which it is deployed. Furthermore, we suggest to narrow the perspective to a single domain, as the effect of a technology is context-dependent. To put this vision into practice, we interview 30 cross-disciplinary experts in the representative domain of rural areas, and we analyse the transcripts to identify common themes. 
%As a result, we provide a classification of technical solutions in rural areas---e.g., automatic milking systems, sensors for grain purity measurement, drones for farm monitoring---and associated drivers, barriers and positive or negative impacts. 
As a result, we provide drivers, barriers and positive or negative impacts associated to the introduction of novel technical solutions in rural areas.
%---e.g., automatic milking systems, sensors for grain purity measurement, drones for farm monitoring. 
This RE-relevant information could hardly be identified if interested stakeholders were interviewed before the development of a single specific system. This paper contributes to the literature with a fresh perspective on sustainability requirements, and with a domain-specific framework grounded on experts' opinions. The conceptual framework resulting from our analysis can be used as a reference baseline for requirements elicitation endeavours in rural areas that need to account for sustainability concerns.    

\end{abstract}

% keywords can be removed
\keywords{Software Engineering \and
Requirements Engineering \and Sustainability Requirements \and
Interviews \and Digitalisation \and Empirical Study}

\section{Introduction}
\label{sec:introduction}

%\hl{Ale: TODO, look into value based design, keynote at REFSQ,} \url{https://link.springer.com/article/10.1007/s00766-017-0273-y}
%\url{https://design-justice.pubpub.org}
%\url{https://journals.sagepub.com/doi/pdf/10.1177/0162243917714869}
%\url{https://evidlab.umd.edu}

Sustainability in system engineering has traditionally been interpreted as the ability of a system to evolve and be maintained in a cost-effective way,  while managing technical debt~\citep{koziolek2011sustainability,becker2015sustainability,kruchten2012technical,li2015systematic,penzenstadler2012sustainability,condori2018characterizing}. This vision, which focuses only on the \textit{technical} side of sustainability, has been criticized by the Karlskrona Manifesto~\citep{becker2015sustainability}, edited by a group of software engineering researchers to raise awareness on the relationship of Information and Communications Technology (ICT) solutions with ecological and social systems. The manifesto calls for a more systemic view of sustainability during system design, and identifies requirements engineering (RE) as the key area where system-level thinking can be applied to escape the trap of solutionism~\citep{easterbrook2014computational}, and broaden the perspective to reason on potential effects of technological change from the social, ecologic and economic viewpoints. 
%When transforming an existing  environment through the introduction of an ICT system, traditional RE approaches normally focus on the analysis of existing processes, stakeholders' needs, and social relations~\cite{eric2011social,pohl2010requirements,van2009requirements}. While this can guide the engineering of suitable solutions 
%that take into account costs, benefits, budget and time within a
%for a short-term perspective, it is not sufficient to guarantee that sustainability concerns are addressed in the long run~\cite{becker2015sustainability,becker2015requirements}. As clearly stated by Becker et al.~\cite{becker2015sustainability}, ``\textit{the system the customer wants and the system that should be built are quite different}''. Design choices may privilege some stakeholders while marginalizing others, and may not consider silent stakeholders, such as the natural ecosystem, animals, and future generations. 

The call to arms of the Karlskrona Manifesto triggered research around the notion of \textit{sustainability requirements}~\citep{venters2017characterising,lago2015framing,mahaux2011discovering,condori2018characterizing,chitchyan2016sustainability}. These are intended as quality goals that a system shall fulfill to provide long-term benefits for its environment and members therein, while minimising damage for other members and the environment as a whole~\citep{venters2017characterising}\footnote{As pointed out, among others, by~\cite{venters2017characterising}, the concept of \textit{sustainability requirement} is not well defined in the literature. Here we provide an intuitive idea to clarify what is the topic of discussion, without any ambition for formality or completeness.}. Different RE approaches have been proposed to elicit this particular type of requirements. Part of them focus on energy-management aspects~\citep{calero2015green}, and use different combinations of RE practices---prototyping, design thinking, goal modelling,  \textit{etc.}---specifically tailored to elicit requirements concerning the energy-efficiency of the system~\citep{ferrario2016values,mahaux2011discovering,KERN2018199}. Others take a domain-agnostic perspective, and propose general sets of sustainability requirements patterns~\citep{roher2013sustainability}, interview scripts~\citep{duboc2020requirements}, as well as guidelines to rethink the software process considering sustainability as a main concern~\citep{seyff2018tailoring,SAPUTRI2021106407,lami2012measuring,bozzelli2013systematic}. 
%with the usage of design thinking and prototyping~\cite{ferrario2016values}, or combinations of traditional RE practices, including the Volere template\footnote{Volere Requirements Template: \protect\url{http://www.volere.co.uk}} for stakeholder analysis, workshops, goal modelling and use case analysis~\cite{mahaux2011discovering}, as well as literature reviews for the elicitation of energy-efficiency assessment criteria~\cite{KERN2018199}. 
%Others take a domain-agnostic perspective, and propose sets of sustainability requirements patterns~\cite{roher2013sustainability}, interview scripts to be adopted when eliciting sustainability requirements for a specific system~\cite{duboc2020requirements}, strategies for negotiation~\cite{seyff2018tailoring} and guidelines, with associated assessment metrics, to rethink the software engineering process taking sustainability into account~\cite{SAPUTRI2021106407,lami2012measuring,bozzelli2013systematic}. 
Despite these efforts, the mapping study by~\cite{garcia2018interactions} on sustainability and software product quality highlights a limitation in the \textit{scope} of the effects that are  considered by the majority of the studies in the field. While the proposed methodologies go beyond the immediate impacts, and consider the so-called second-order effects---i.e., potential changes in the behaviour of  individual users---most of them do not account for \mbox{third-order} effects, related to the societal and long-term influence of the technological transformation. 

%To address this problem, Duboc et al.~\cite{duboc2020requirements} suggest to ask particular questions specifically oriented to elicit sustainability requirements. The questions are designed for stakeholders that will be involved in system development or usage, and could have personal interests to purposely neglect sustainability aspects.  Furthermore, the approach is domain-independent, and one needs to tailor the questions for the specific project and context.  

This paper posits that, to address existing limitations in terms of sustainability requirements elicitation, going beyond the concept of system requirements and stakeholders' goals is necessary, also raising the degree of abstraction. To this end, we propose to analyse three core concepts, namely \textit{drivers}, \textit{barriers} and potential \textit{impacts} associated to the introduction of novel ICT solutions in a certain socio-physical domain. 
%, from the viewpoint of a cross-disciplinary group of experts. 
These three concepts incorporate traditional stakeholders' goals among the drivers, but also account for other components that are relevant for sustainability, and do not currently have a prominent place in RE. 
We focus on \textit{rural areas}---including rural communities, agriculture and forestry---as this is a representative yet diversified domain that is facing deep technological transformations~\citep{trendov2019digital,doerr2018reinrural,bacco2019digitisation}. Domain specificity is a relevant aspect, as~\cite{penzenstadler2012sustainability} already observed that sustainability should be addressed with domain-dependent lenses.

To practice our vision and elicit information for the three core concepts, we perform a set of 30 semi-structured interviews with experts across the European Union (EU), which were recruited in the context of the Horizon 2020 DESIRA Project (Digitisation: Economic and Social Impact in Rural Areas) \footnote{\url{https://desira2020.eu}}. The experts have diversified knowledge about a wide range of ICT solutions applied in rural areas---e.g., precision agriculture, blockchain-based tracking, and automated milking systems (AMSs). They are selected as they are experts on families of systems in the domain, and can therefore provide an informed opinion, with the right high-level perspective that gives a (filtered) voice to multiple stakeholders. Furthermore, the experts are free from the conflicts of interest that may arise if stakeholders involved in a specific project were interviewed. We perform a thematic analysis of the interview transcripts to identify common categories and provide an expert-based reference framework to be placed before any project-specific requirements elicitation activity in the domain of rural areas. 

Our results show that typical \textit{barriers} for the adoption of ICT solutions are the lack of connectivity in rural areas, but also fear and distrust towards technology. In addition, the cost of technology and regulatory issues, also related to unclear data governance are relevant barriers. Main \textit{drivers} are economic, as  technology can lead to cost reduction, but also ecological and institutional, since  technology can improve monitoring as well as accountability. In this regard, regulators can play a crucial role by means of funding programs and norms. \textit{Positive impacts} are the  replacement of repetitive labour and the possibility of exploiting economies of scale. On the other hand, \textit{negative impacts} are the higher dependency from technology as well as the social exclusion of some players that cannot cope with the change, at least not fast enough.

This work contributes with a paradigm shift in the analysis of sustainability requirements, by introducing the concepts of drivers, barriers and impacts associated to the adoption of technological solutions. Furthermore, our themes represent a reliable snapshot of the state of affairs in rural areas, and can be taken as reference for the development of socio-technical systems in this domain. In addition, by raising the level of abstraction of RE, our work paves the basis for further integration between sustainable development and value-based software engineering~\citep{mougouei2018operationalizing,ferrario2016values,newman2015role}.

The remainder of the paper is structured as follows. In Section \ref{sec:background}, we analyse related works especially in the field of RE-based sustainability, and shed some lights on digitisation processes in rural areas. The research design is presented in Section \ref{sec:design}, detailing the interview scripts, the selection of the interviewed experts, and their expertise and provenance. Section \ref{sec:results} analyses the collected results, and categories them into socio-cultural, technical, economic, and regulatory related themes, providing insights on drivers, barriers, and potential impacts. Section \ref{sec:discussion} 
The conclusions are in Section \ref{sec:conclusion}.
%TEX root = ../paper.tex
\section{Background and Related Work}
\label{sec:background}

In the following we first introduce the EU H2020 DESIRA project, in which this work is conducted. Then, we provide an overview of related work on RE for sustainability requirements, and we highlight our contribution with respect to the literature. 
%Finally, we present the elements of the conceptual framework that drive the current research. 

\subsection{The H2020 DESIRA Project}
The paradigm of cyber-physical systems \citep{wolf2009cyber} is often referred to as a model to describe how complex systems interact with the physical world, integrating computation and physical processes. Depending on the context, the cyber and physical spaces can be intertwined with the social space \citep{lace2018required}, giving birth to the concept of socio-cyber-physical systems, a paradigm in which humans are at the very center, as opposed to cyber-physical systems that revolve around computation and physical processes. The \textit{socio-cyber-physical} paradigm is the core of DESIRA (Digitisation: Economic and Social Impacts in Rural Areas)~\citep{desira2020}, a four-year H2020 EU project started in June 2019, which focuses its attention on the digitalization of rural areas, including agriculture, forestry and rural communities. 
The analysis conducted within DESIRA covers both the past and the present, and also aims at developing future scenarios in which the impacts of digital technology can be defined as \textit{game changing}. A digital game changer can be defined as a disruptive digital technology introduced or adopted in a context\footnote{See a conceptual briefing on Digital Game Changers at: \protect\url{https://desira2020.eu/wp-content/uploads/2020/11/Briefing_Digital-Game-Changers.pdf}}.
%For instance, the introduction and widespread use of AMSs in the last \hl{10?} years has been dubbed as a game-changing event in the case of livestock. 
%In fact, AMSs provide a substantial support to farmers, freeing time for other tasks; furthermore, each animal can approach the machine whenever needed, which creates positive effects on its welfare; finally, the use of disinfectant solutions strongly reduces the risk of infections, as well as improves the quality of the milk. Thus, their use has disrupted dairy farming procedures, with impacts at economic level (e.g. lower infection rate and higher milk quality), social level (more free time for other tasks), and animal welfare (each animal decides when to be milked, but fewer contacts with owners).
The socio-economic impacts of potential digital game changers are discussed in twenty Living Labs\footnote{The Living Labs can be seen on the DESIRA website: \protect\url{desira2020.eu}.} all across Europe, each around its own focal question that embodies a crucial need or desire in a geographical area. For instance, how to digitally trace wood over the entire process lifecycle in a way that is economically and bureaucratically sustainable for forest owners/managers, but also facilitating the work of both certification and control entities. 

In order to identify and assess the socio-economic impacts of digitalization in rural areas, the DESIRA project will put forward both conceptual and analytical tools, to be used in the assessment of the past and present situation in the 20 Living Labs. Furthermore, the Living Labs will also perform the so-called \textit{scenario workshops} to explore different future scenarios with respect to game-changing events, such as the adoption of digital technologies that have the potential to reshape rural areas. The Living Labs will also co-design novel digital solutions tailored on the specificity of rural areas. The co-design will be carried out in the so-called \textit{use case workshops}, involving relevant stakeholders from different sectors as in the scenario workshops. The use cases will be put forward by the DESIRA project as instances of \mbox{high-level} technological solutions for which a previous discussion around drivers, barriers, and impacts has been carried out, thus lowering the risks of unintended effects due to digitalization \citep{scholz2018unintended}. The methodology will revolve around the \textit{Responsible Research and Innovation}\footnote{\protect\url{https://ec.europa.eu/programmes/horizon2020/en/h2020-section/responsible-research-innovation}} approach, a framework to guide the development and introduction of new technologies in a manner that identifies, accommodates, and responds to and addresses societal concerns.

In this work, we focus on the creation of a baseline framework of drivers, barriers and impacts of digitalization in rural areas, based on experts' interviews. This is the starting point of the DESIRA analysis. The framework will be further specialised, considering the specific contexts of the Living Labs as novel relevant elements will emerge along with the scenario and use case workshops. 

\subsection{Sustainability in Requirements Engineering}
When transforming an existing context through the introduction of an ICT system, traditional RE approaches normally focus on the analysis of existing processes, stakeholders' needs, and social relations~\citep{eric2011social,pohl2010requirements,van2009requirements}. While this can guide the engineering of suitable solutions 
that take into account costs, benefits, budget and time within a
for a short-term perspective, it is not sufficient to guarantee that sustainability concerns are addressed in the long run~\citep{becker2015sustainability,becker2015requirements}. As clearly stated by~\cite{becker2015sustainability}, ``\textit{the system the customer wants and the system that \underline{should} be built are quite different}''. Design choices may privilege some stakeholders while marginalizing others, and may not consider silent stakeholders, such as the natural ecosystem, animals, and future generations. It is therefore important to provide means to reason on \textit{sustainability requirements}~\citep{venters2017characterising,lago2015framing,mahaux2011discovering,condori2018characterizing,chitchyan2016sustainability,volkov2018smart,penzenstadler2014safety}, intended here as quality goals that a system shall fulfill to provide long-term benefits for its environment and members therein, while minimising damage for other members and the environment as a whole. 

In recent years, several works have been conducted to address the challenge of eliciting, analysing and satisfying sustainability requirements. Part of the work focuses on experimenting and tailoring RE methods. Others are oriented to surveying the field and provide general frameworks. 
In the following, we summarise representative contributions in the two groups.

\subsubsection{RE Methods for Sustainability} Research in RE and sustainability dates back to the late '00, with the seminal work of ~\cite{cabot2009integrating}. The authors propose to use the well-known $i^*$ goal modelling framework to represents the sustainability effect of each business or design alternative. Sustainability is defined as a \textit{softgoal} (i.e., a nonfunctional/quality requirement) and is further decomposed into subgoals, such as reuse, recycle, etc. to build a reference taxonomy. The approach is applied to a preliminary case study. The main research challenges observed are related to the absence of standard definitions of sustainability concepts and metrics, and scalability issues of the $i^*$ modelling language.~\cite{mussbacher2014goal} introduce goal-oriented engineering for sustainability, and uses the Goal-oriented Requirements Language (GLR), extended with the notion of time to account for measurable aspects relate to this variable and its relation to sustainability.  In another work, Roher and Richardson propose to use a recommender system for sustainability requirements, so to enable reuse of requirements archetypes~\citep{roher2013proposed}. The same authors further develop the concept of archetypes into \textit{sustainability requirements patterns}~\citep{roher2013sustainability}, and derive three main patters related to resource consumption from the analysis of existing documents. 

\cite{mahaux2011discovering} take a more empirical perspective, with an experience report oriented to reflect on the process of discovering sustainability requirements. They use a combinations of traditional RE practices, including the Volere template\footnote{Volere Requirements Template: \protect\url{http://www.volere.co.uk}} for stakeholder analysis, workshops, goal modelling with KAOS~\citep{van2009requirements} and use case analysis. The paper observes that sustainability requirements are qualities that can be analysed using traditional techniques. On the other hand, it also highlights that specific checklists need to be defined, and, most of all, a sustainability specialist need to be involved in the RE process.

\cite{brito2018} combine aspect-oriented requirements analysis with the hybrid assessment method, an approach for multi-criteria decision making. They define a meta-model to represent sustainability concerns, which includes the potential \textit{effect} of a certain requirement, a notion similar to the one of impact that we consider in our paper. The approach is experimented in a case study with unmanned aerial vehicles (UAV) for agriculture. 
%Penzestadler \textit{et al.}~\cite{penzenstadler2018} focus on analysing sustainability of existing systems, and propose to use leverage points, defined as locations where a small change in one aspect can have a positive systemic effect. 

\cite{seyff2018tailoring} tailor the Win Win negotiation process to consider the impact of requirements on sustainability. The approach is applied on an industrial case study involving an ERP system vendor. Though the experience was considered successful, discussion on the impact of requirements was hampered by a lack of information to anticipate long-term effects, which lead to participants having different, and uncertain, opinions. Specific to the context of rural areas, ~\cite{doerr2018reinrural} present an RE framework to assess and derive new RE methods for social contexts. The authors have previously experimented with design thinking within Living Labs---a paradigm also used in DESIRA---, demonstrating the effectiveness of the approach. Based on the experience, their framework highlights the need to consider different RE dimensions, including the attitude of people towards IT systems, as well as the impact of the technology.  
%They argue that, in a social context such as rural areas, the involvement of \textit{future} users is of the highest importance. 

In a recent work,~\cite{duboc2020requirements} present an RE framework to facilitate the elicitation of sustainability-related requirements according to the five dimensions identified by ~\cite{becker2015sustainability},  
%based on Goodland~\cite{goodland1995concept},
namely \textit{individual}, \textit{social}, \textit{technical}, \textit{economic} and \textit{environmental}. The framework consists of a set of questions to be asked to stakeholders during interviews or workshops. The questions are  specifically oriented to facilitate reasoning on the short and long term impacts of the deployment and usage of a certain system. The framework also includes a diagrammatic notation to graphically support a coherent analysis of the relationship between the different types of impacts across the dimensions. 

Saputri \textit{et al.}~\citep{SAPUTRI2021106407,saputri2020addressing} propose a complete framework, with guidelines to elicit and assess sustainability requirements and metrics. They use the Goal-Question-Metric (GQM) approach and partition requirements into the dimensions used by~\cite{duboc2020requirements}. The approach is applied on different case studies, showing that the guidelines provided facilitate the identification of sustainability requirements. On the other hand, difficulties were encountered in eliciting domain-specific sustainability metrics.

%However, traditional ways of requirements elicitation are not appropriate because e.g. the potential users lack IT skills, have never participated in the development of IT Systems, or have even negative perceptions against IT. The proposed framework provides characterizations of end users in their social context as well as of RE methods.

\subsubsection{Surveys on RE for Sustainability} Based on previous works proposing RE solutions, and considering also surveys in the broader area of software engineering for sustainability~\citep{penzenstadler2012sustainability,penzenstadler2014systematic}, 
~\cite{chitchyan2015evidencing} gives an overview of techniques that can be applied to support sustainability in each RE-relevant phase (feasibility study, stakeholder analysis, elicitation, documentation). 
On a similar note,~\cite{garcia2018interactions} present a mapping study on sustainability and software product quality, noticing that this is a particularly lively area of research, but still at its exploratory stage, with works that are mostly focused on the development of energy-saving solutions, which are only one of the multiple facets of sustainability. In another contribution~\citep{mireles2017classification}, the same authors focus on surveying RE methodologies for sustainability, and notice that, while several approaches have been presented and experimented, there is limited knowledge on how to \textit{assess} the achievement of sustainability requirements. 
%Furthermore, the work highlights a limitation on the \textit{scope} of the effects that are considered by current studies. While the proposed methodologies go beyond the immediate impacts, and consider the so-called second-order effects---i.e., potential changes in the behaviour of individual users---they do not consider third-order effects, related to the societal and \mbox{long-term} impacts of the technological transformation.

While these work mostly focus on gathering data from the literature,~\cite{chitchyan2016sustainability} look more into practice, performing an interview study with RE professionals to identify their viewpoints on sustainability requirements. Among the different aspects, the subjects generally complained about the absence of a clear development methodology to support sustainability in their companies, and the lack of support for engineers in understanding sustainability issues. Similarly,~\cite{condori2018characterizing} perform an online survey with different software professionals to identify how different quality requirements, framed according to the ISO/IEC 25010:2011 Quality model~\citep{25010:2011}, contribute to sustainability. Building on a previous work from~\cite{lago2015framing}, they analyse the responses according to four sustainability dimensions, namely: \textit{social}, \textit{technical}, \textit{economic} and \textit{environmental}. The results show that the different dimensions are intertwined, as a type of requirement can address multiple dimensions at once. For example, availability and efficiency requirements address the technical dimension, but are also strongly related to the environmental and economic ones.

\subsubsection{Contribution} Our work falls into the group of works concerned with surveys about sustainability for RE (e.g., ~\cite{condori2018characterizing,chitchyan2016sustainability,garcia2018interactions}). With respect to previous work in this group, this is the first one that does not consider RE practitioners as subjects. Instead, we collect the viewpoints of sustainability experts. In this sense, our work complements existing literature in RE by giving voice to those experts whose role is considered to be extremely valuable by previous authors~\citep{mahaux2011discovering,chitchyan2016sustainability,seyff2018tailoring,Fuentes2010HumanContextInRE,Leimeister2014Digitalservices}. Although a full-fledged RE methodology is not within the scope of this paper, our work also aims to contribute with a novel view on sustainability requirements, by introducing the concepts of drivers, barriers and impacts. This view, which is summarised in the following, can act as a reference framework to develop further methodologies to support sustainability in RE.

\section{Reference Conceptual Framework}
\label{sec:framework}

Traditionally, RE has revolved around the concepts of stakeholders, actors, goals (or functional requirements), softgoals (or quality requirements, or nonfunctional requirements), domain assumptions, and specifications~\citep{van2009requirements,pohl2010requirements}. Sustainability requirements are generally considered by the literature as a form of \textit{softgoal}~\citep{mahaux2011discovering,volkov2018smart,venters2017characterising,penzenstadler2014safety}. They require to reason on the impact that a system can have on the context in which it is deployed in terms of second-order effects (e.g., indirect changes in user behaviour), and third-order ones (e.g., societal and long-term influence due to rebound effects)~\citep{penzenstadler2014safety}. According to~\cite{garcia2018interactions}, current approaches tend to be insufficient in addressing the latter types of effects, which account for elements that are related to culture, society, economy, politics and other collective aspects that characterise a socio-cyber-physical context. These elements can contribute to facilitating or hindering the acceptance of a certain technological change, and they should be explicitly considered as main focal points when reasoning on  sustainability requirements. Furthermore, in line with other researchers~\citep{mahaux2011discovering,chitchyan2016sustainability,seyff2018tailoring,Fuentes2010HumanContextInRE,Leimeister2014Digitalservices}, we argue that specific experts need to be involved when reasoning around sustainability. 

%We thus propose to account for these elements by taking a higher-level perspective, and explicitly introducing the concepts of \textit{drivers}, \textit{barriers} and \textit{impacts} that a certain digital system can have in the context in which it is deployed. 

\begin{figure}[h]
\includegraphics[width=0.8\textwidth]{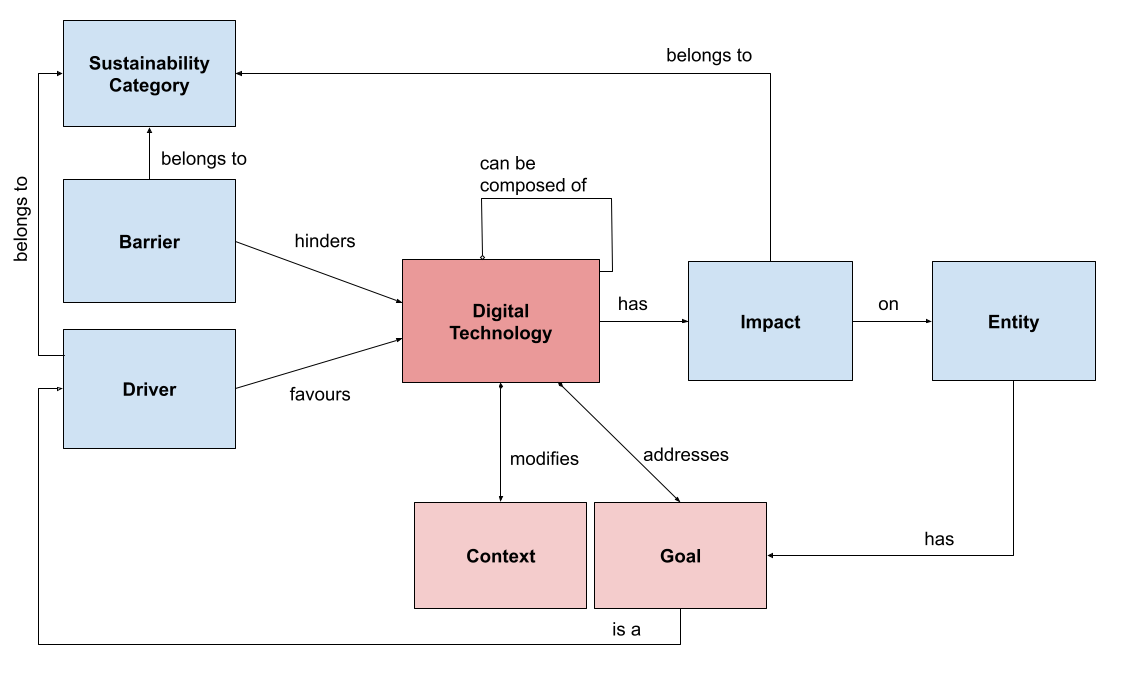}
\centering
\caption{Meta-model of the proposed reference framework. Light-blue elements are the topic of this paper.}
\label{fig:diagram}
\end{figure}

We thus propose to adopt the high-level concepts of drivers, barriers and impacts related to the introduction of a certain technological application in an existing socio-cyber-physical context\footnote{We refrain from the usage of the term \textit{environment}, which is more commonly used in RE, as in the term is reserved to refer to ecosystems}. These concepts are analysed building upon the different sustainability categories adopted in previous studies~\citep{lago2015framing,duboc2020requirements}. Furthermore, we propose to elicit information for these aspects from selected sustainability experts in a given domain. Fig.~\ref{fig:diagram} reports an informal meta-model that summarises our vision. In the following, we describe the main concepts and their relations.

A \textbf{Digital Technology} represents a family of digital systems, or composition thereof, which aims at satisfying or satisficing a given set of hard- and soft- \textbf{Goals}, and in doing so it modifies an existing socio-cyber-physical \textbf{Context}. For example, a vegetation monitoring technology based on hyper-spectral cameras and signal processing can have the goals of monitoring the field and ensure grain quality. The technology socially and physically modifies a context made of farmers (e.g., by introducing technological experts) and fields (e.g., by introducing cameras carried by drones).    

The introduction of the technology in the context is favoured by \textbf{Drivers} and hindered by \textbf{Barriers}, and has certain \textbf{Imapacts} on existing  \textbf{Entities}.

\textbf{Drivers} include goals of some stakeholders, for example the need to improve wheat quality required by farmers, but also other higher-level aspects, for example the funding from institutions to support  specific technologies. Similarly, \textbf{Barriers} include obstacles in KAOS terms~\citep{van2009requirements}, intended as elements preventing the achievement of a specific goal, but also more structural impediments that hamper the introduction of the digital technology as a whole in the given context. For example, the difficulty of farmers in interacting with the novel technology, or the regulatory problems related to the usage of drones. 
%The Impact is intended as a mid- to long- term effect on a certain entity. 
The concept of \textbf{Impact} is analogous to that already considered, among others, by~\cite{brito2018} and by~\cite{seyff2018tailoring}, and is intended as the expected effect that the digital technology can have from a sustainability standpoint, and thus in mid- to long-term. The impact can be positive, as, e.g., reduction of manual labour, but also negative, for example due to the exclusion of small farmers that cannot afford the technology. An \textbf{Entity}, instead, includes actors, stakeholders, and any party that is indirectly impacted by the technology without voluntarily interacting with it or taking part to the decision process that leads to its deployment, such as the environment, the animals, or the community as a whole.  
Drivers, barriers and impacts are partitioned into different sustainability categories. Base categories, or dimensions, are \textit{social}, \textit{technical}, \textit{economic}, \textit{environmental}, and \textit{individual}, as in~\cite{duboc2020requirements},~\cite{lago2015framing}\footnote{~\cite{lago2015framing} does not explicitly include the \textit{individual} dimension.} and other authors~\citep{becker2015sustainability,goodland1995concept}, but they can be extended or renamed based on the specific data gathered in the given domain. 

In the following, we specialise the framework for the domain of rural areas by interviewing experts in sustainability in relation to the introduction of digital technology, or \textit{digitalisation}, for short. We focus on the main elements of drivers, barriers and impacts, and we relate them with sustainability categories, and impacted entities in the specific domain. The resulting framework aims to define a knowledge base that can be useful to RE endeavours in rural areas concerned with the development of novel systems that need to take sustainability into account. 

%This reference framework can in principle be applied to any domain. Drivers, barriers and impacts need to be elicited from different experts in the domain. Based on this elicited information, specific applications can then be developed in a certain context. In the paper, we focus on eliciting information related to the light-blue elements of the diagram. This information can be used to guide the development of specific technologies, or digital system design, taking long-term sustainability concerns into account. 

%Given a specific technology, or digital system design, with defined goals and context, this information can be used to enrich the RE process in terms of sustainability requirements.

%We believe that similar endeavours should be conducted also in other context, including IT companies, to facilitate the identification 

%The analysis of the interviews is based on the reference concepts of drivers, barriers and impacts, which have been further refined into categories and themes. Stemming from the analysis, we now summarise the relationships between the different ingredients. Fig.~\ref{fig:diagram} represents the meta-model of the derived framework.
\section{Research Design}
\label{sec:design}
The present study can be regarded as a \textit{judgment study}~\citep{stol2018abc}, which is a form of in-depth survey involving selected experts on a certain topic of interest---in our case digitalisation in rural areas. We use semi-structured interviews as data collection technique. The study is carried out by first selecting a set of representative experts as participants, and then by interviewing them according to pre-defined interview scripts designed to collect their opinion, and tailored to their profiles. The interviews are transcribed, translated into English, and their content is analysed through open coding followed by axial coding, to produce a coherent and complete view of the topic of interest based on the collected opinions. The study is \textit{exploratory} and \textit{descriptive} in nature, as it is oriented to provide a first overview requirements-relevant aspects related to digitalisation in rural areas. Therefore, our goal is not to explain the observed phenomena, but rather to provide a descriptive reference framework of the current digitalisation landscape, to facilitate the approaching of requirements engineers to the domain. 

\subsection{Research Questions}
\label{sec:resquestions}

The overall objective of the study is to identify barriers, drivers and impacts of digitalisation in rural areas. This objective is decomposed into the following four research questions (RQs). 

%\textbf{RQ1:} \textit{What are the main technologies that play a role in the digitalisation of rural areas?} The question aims to identify the scope of the considered domain in terms of the object(s) of discussion, namely the technologies that are used or can be used in the future for the digitalisation of rural areas. Although each single technology can be in principle associated to specific drivers, barriers and impacts, our goal is not to provide a fine-grained one-to-one mapping, but rather to provide a comprehensive view of what digitalisation means for rural areas. 

\textbf{RQ1:} \textit{What are the barriers hindering digitalisation in rural areas?} The question aims to identify what are the elements that inherently hamper the introduction of digital technologies. Barriers are intended in a broad sense, without a specific definition, so that economic, technical, social and other aspects can emerge without a strict focus on one of the facets. 

\textbf{RQ2:} \textit{What are the drivers facilitating digitalisation in rural areas?} The question aims to identify the elements that push towards digital transformation. As for barriers, drivers are intended in a broad sense and include also what can be regarded as \textit{goals} pursued by certain stakeholders. Therefore, one can identify drivers that cannot be fully controlled and exist without a specific rural actors pushing for them (e.g., the decreasing cost of technology) as well as goals of well-defined actors (e.g., need of farmers for better control of the production). 

\textbf{RQ3:} \textit{What is the potential impact of digitalisation in rural areas?} The question aims to describe positive and negative consequences that rural socio-technical systems can experience when increasing the strength of their technical side with the introduction of digital technologies. The objective of the question is to reflect on middle- to long-term impacts that any project with a strong digital component could have in this domain. 

\subsection{Study Participants Selection}
\label{sec:participants}
Participants of the study were selected by the authors based on opportunistic sampling. The goal was to involve experts that: (a) could cover the main sub-domains of rural areas, namely agriculture, forestry and rural communities; (b) covered ICT and social-science background; (c) could be representative of different geographical areas of the EU. The participants to the DESIRA project, who have interdisciplinary backgrounds including ICT, social science and agriculture, contacted specific subjects in their fields that were considered as reliable experts due to their professional position and their publicly recognised active role in the theme of digitalisation for rural areas. The selected experts do not have a role in the DESIRA project. Table \ref{tab:experts} lists the selected participants together with their reference subdomain, nationality, main expertise and gender. In Table~\ref{tab:technologies}, instead, we list the reference technologies considered, to give an indication of what is the technological scope covered by the experts. 

\begin{sidewaystable}
\begin{scriptsize}
     \centering
    \begin{tabular}{c|cccc}
    \hline
    \textbf{ID} & \textbf{Sub-domain} & \textbf{Geographical Area} & \textbf{Main Expertise and Role} & \textbf{Gender} \\ \hline \hline
    1 &  agriculture & France & academic, ICT researcher & M \\ \hline
    2 & agriculture & France & support for policy making & M \\ \hline
    3 & agriculture & France & operation director in private company & M \\ \hline
    4 & agriculture & France & instructor and consultant for agricultural cooperatives & M \\ \hline
    5 & agriculture & Finland & academic, automation technology in farms & M \\  \hline
    6  & agriculture & Belgium & academic, ICT researcher & M \\ \hline
    7  &  agriculture & Greece & academic, ICT researcher & M \\ \hline
    8  & agriculture & Greece & social science, researcher & M \\ \hline
    9  & agriculture & Switzerland & agricultural research & M \\ \hline
    10  &  agriculture & Latvia & consultant, researcher & F \\ \hline
    11   & agriculture & Latvia & academic, ICT researcher & M \\ \hline
    12  &  agriculture & Germany & state research center for agriculture & M \\ \hline
    13  & agriculture & UK & head of farms networks & M \\ \hline
    14  &  agriculture & Hungary & farm manager & M \\ \hline
    15  & agriculture & Italy & agronomist, researcher & M \\ \hline
    16  & agriculture, rural communities & Finland & ICT project manager, researcher & M \\ \hline
    17  &  agriculture, rural communities & Spain & rural development, support for policy making & F \\ \hline
    18  & agriculture, rural communities & France & sociologist, focus on rural areas & F \\ \hline
    19  &  rural communities & France & advisor, entrepreneur & F \\ \hline
    20  &  rural communities & Netherlands & academic, rural and community development & F \\ \hline
    21  &  rural communities & Netherlands & researcher in ethics, impacts of innovation & F \\ \hline
    22  &  rural communities & Spain & manager of a natural protected area & M \\ \hline
    23   &  rural communities & Belgium & policy expert & M \\ \hline
    24  &  rural communities & Germany & academic, ICT researcher & M \\ \hline
    25  &  rural communities & Poland & agriculture and food economics & M \\ \hline
    
    26  &  rural communities, forestry & Greece & academic & F \\ \hline
      
    27  & forestry & Italy & manager of non-profit consortium, forest engineer & M \\ \hline
    28  &  forestry & Italy & startup founder, renewable sources & M \\ \hline
    29  &  forestry & Austria & education, training, research & M \\ \hline
    30 &  forestry & Spain & head technical team of environmental information network & M \\ \hline

    \end{tabular}
    \caption{Interviewed experts.}
    \label{tab:experts}
\end{scriptsize}    
\end{sidewaystable}
\begin{table}
\begin{scriptsize}
     \centering
 \begin{tabular}{c|p{0.9\textwidth}}
    \hline 
\textbf{ID} & \textbf{Technologies} \\ \hline \hline
1 & automatic miking systems, sensors, agri-voltaic \\ \hline
2 & service dematerialization \\ \hline
3 & artificial intelligence, blockchain \\ \hline
4 & sensors, GPS, blockchain, precision agriculture, agriculture robots \\ \hline
5 & automatic miking systems, assisted driving for agriculture, cellular agriculture, controlled environment agriculture (CEA) \\ \hline
6 & precision agriculture, automatic milking systems, sensors, cameras, satellite images \\ \hline
7 & digital communication \\ \hline
8 & animal tracking, vehicle tracking, UAV/drones, satellite images, cameras, precision farming \\ \hline
9 & sensors, imagine, machine learning \\ \hline
10 & GPS, tractor tracking, precision agriculture, precision farming, drones \\ \hline
11 & automatic milking robots, precision farming \\ \hline
12 & assisted driving for agriculture, agriculture robots \\ \hline
13 & earth observation, satellites, 3D imaging for forests, Web-GIS, habitat modelling, pest prediction, infestation prediction \\ \hline
14 & sensors, precision farming \\ \hline
15 & Web-GIS, remote sensig, pest monitoring and prediction, machine learning, speech recognition \\ \hline
16 & IoT, computer vision, 5G, drones, energy monitoring, artificial intelligence, precision agriculture, sensors, visual scanners \\ \hline
17 & open data, IoT, artificial intelligence, satellite images, Web-GIS, drones, sensors, robots \\ \hline
18 & data sharing, social networking \\ \hline
19 & semantic web, remote consultants \\ \hline
20 & drones, precision agriculture, satellites, smart tractors, sensors, automatic driving for agriculture \\ \hline
21 & social network, digital communication \\ \hline
22 & automated driving for agriculture, drones, cameras, weather sensors, soil analysis, precision agriculture \\ \hline
23 & remote health, distant education, online marketing \\ \hline
24 & cloud technologies, IoT, mobile apps, virtual reality, augmented reality, artificial intelligence \\ \hline
25 & precision farming, Web-GIS, data mining \\ \hline
26 & open data, precision agriculture, wildfire prediction \\ \hline
27 & blockchain, QR code, RFID \\ \hline
28 & QR code, blockchain, RFID, sensors, wildfire prediction \\ \hline
29 & soil monitoring, livestock monitoring, crop monitoring, GPS, sensing, satellite images, UAV/drones, artificial intelligence, precision agriculture \\ \hline
30 & satellite images, sensors, UAV/drones, meteorological data, IoT \\ \hline
    \end{tabular}
    \caption{Technologies considered by the experts.}
    \label{tab:technologies}
\end{scriptsize}    
\end{table}
\begin{sidewaystable}
\begin{scriptsize}

    \centering
    \begin{tabular}{c|c|c}
     & \textbf{ICT Expert} & \textbf{Socio-economic Expert} \\ \hline \hline                                                                                                                         
    \textbf{Q1} & \multicolumn{2}{c}{Which are the DTs you deal with or encounter most commonly in your work? What are their main uses in the three domains?} \\ \hline \textbf{Q2} & Which is a plausible tomorrow’s use of the DTs you cited in Q1? & What are the socio-economic impacts of the DTs you cited in Q1? \\ \hline       \textbf{Q3} & \begin{tabular}[c]{@{}c@{}} Can you provide some examples of uses of DTs / new developments\\you are participating to/aware of? Do you think\\those developments have the potential to be game changers?\end{tabular} & \begin{tabular}[c]{@{}c@{}}What do you consider as drivers for the adoptions of DTs\\in the three domains?\end{tabular} \\ \hline                             
    \textbf{Q4} & \begin{tabular}[c]{@{}c@{}}Which are the positive and negative impacts of technological\\advancement on SMEs, workers, and other actors,\\especially considering cases you have been involved into?\end{tabular} & \begin{tabular}[c]{@{}c@{}}What do you consider as barriers for the adoptions of DTs\\in the three domains?\end{tabular} \\ \hline
    \textbf{Q5} & \begin{tabular}[c]{@{}c@{}}What do you consider as drivers and barriers for the adoption\\ of DTs in the three domains?\end{tabular} & \begin{tabular}[c]{@{}c@{}}How new and deeper reflections / methodologies to assess\\the impacts of technology could help you in your work?\end{tabular} \\ \hline
    \textbf{Q6} & \multicolumn{2}{c}{Have you already been involved in any activities to assess the socio-economic impacts of DTs?} \\ \hline
    \end{tabular}
    \caption{Interview scripts covering digital technologies (DTs). Q1 and Q6 are common for both profiles.}
    \label{tab:interview-scripts}

\end{scriptsize}    
\end{sidewaystable}

\subsection{Data Collection and Analysis}
\label{sec:datacollectionanalysis}
To collect data, we first defined a set of interview scripts to guide the interviews and then we performed a form of thematic analysis~\citep{vaismoradi2013content,auerbach2003qualitative}, by means of open coding followed by axial coding.  

\paragraph{Interview Scripts and Delivery}
The selected subjects have an interdisciplinary background, but broadly belong to two groups: social-science experts, and ICT experts. Therefore, we defined two main interview scripts, one for each group. The questions for the two groups are reported in Table~\ref{tab:interview-scripts}. Interviews were conducted remotely by the different authors of this paper and by other partners of the consortium, and then transcribed. The transcription was checked by the interviewed subjects for misunderstanding. 

\paragraph{Interview Analysis}
Each interview was initially evaluated by the first author in two cycles. In the first coding cycle, from each interview, he extracted independent paragraphs and coded them based on their content, and following the coding guidelines of~\citep{saldana2021coding} for \textit{descriptive coding} by associating descriptive themes to them. In a second cycle, the themes were aggregated into higher-level sustainability categories, by means of \textit{axial coding}~\citep{saldana2021coding} and leveraging the sustainability dimensions from the literature~\citep{lago2015framing,duboc2020requirements,becker2015sustainability,goodland1995concept}. He used a shared spreadsheet file (a Google sheet) to record themes and categories. From this hierarchical grouping, he produced a set of summary tables that answer the different RQs. The link between data, themes and categories were cross-checked by the third author, who had access to the spreadsheet, and commented for unclear links or theme names, to come to a consolidated output.

\subsection{Threats to Validity}
\label{sec:validity}

Validity of the findings is discussed according to the categories of validity, reliability, and generalisation outlined by~\cite{leung2015validity}.  

\paragraph{Validity} The main requirement for judgment studies is the adequate expertise of the subjects involved, so that the collected opinions are  authoritative and informed ones~\citep{stol2018abc}. The level of expertise of the selected subjects was checked by the DESIRA project consortium, which is formed by multiple institutions that study rural areas from different viewpoints (ICT, economic, legal, etc.), and have an up-to-date vision of relevant voices in the field. To balance the specific background of each subject, two types of script were defined, one for ICT experts and the other for social-science experts. To increase content validity, the scripts were reviewed and piloted within the consortium.  Concerning the completeness of the information collected from each participant with respect to the RQs, we defined interview questions that are derived from the RQs, but are also  sufficiently broad to allow interviewees to freely and completely express their opinions on the discussed topics. A limitation of the study is the reduced number of negative impacts elicited, as the subjects appeared to mostly emphasise positive aspects of digitalisation. Further work within the Living Labs will be conduced with interviews oriented to stress on negative aspects. 
%Therefore, a set of 2 extra interviews were performed, with the same scripts and involving additional subjects, in which the interviewer specifically asked to stress on negative aspects. 
Member checking was adopted to ensure descriptive validity, as the interviewee could review and correct their transcribed interviews. 

\paragraph{Reliability} Different forms of structured procedures were adopted to support triangulation and increase the reliability of the findings: (a) the coding activities  were applied to the whole text of each interview, and all codes and associated interview data were shared in a spreadsheet, to facilitate cross-checking; (b) one researcher performed the coding activity and a second one cross-checked the results with respect to the original data; (c) the resulting findings (i.e., the preliminary versions of the tables reported in Sect.~\ref{sec:results}) were analysed by the other authors of the paper in a meeting in which they could contrast the results with their previous knowledge, and their experience as interviewers. In case some themes emerged that were not familiar to the authors, they were discussed to better consolidate the results with respect to the collected data; (d) excerpts are reported from the interviews that show evidence of the relation between codes and data.

\paragraph{Generalisation}
In our study, experts were selected to have a sufficient coverage of three main dimensions (subdomain, geographical EU area, and background), as reported in Table~\ref{tab:experts}. Therefore, their opinions, and our findings,  mainly reflect their background. In particular, the results are representative for the subdomains of agriculture and rural communities in both southern and northern EU countries, and for forestry, but mostly in southern EU countries. Different results may be obtained if other continents are considered.

\section{Execution and Results}
\label{sec:results}

% \begin{figure*}[h]
% \centering
% \includegraphics[width=\textwidth]{img/diagram.png}
% \caption{Concepts and relationships derived from the study in the form of a simplified class diagram.}
% \label{fig:diagram}
% \end{figure*}

% \cite{sargent2017social} What Is IT for Social Impact?: A Review of Literature and Practices

% \hl{\textbf{NOTES}
% Key concept: identifying requirements of socio-cyber-physical systems requires to take into account also barriers and needs at the social level. Indeed, identifying system requirements is not sufficient, as barriers could exist---at social, political and cultural level---that do not allow the realisation of the requirements. At the same time, it is needed to exploit and control drivers so that changes are facilitated.  

% Finding: Most of the barriers are socio-cultural, while the drivers are environmental or economical.

% From the barriers we can infer new drivers not explicitly mentioned by the stakeholders, as drivers appear to be more limited in number. 

% Drivers are also enablers. Part of the drivers are also needs to be achieved, so the drivers include goals as well.}

Interviews were conducted between May 2020 and February 2021. Results were analysed between October 2020 and April 2021. This section reports the results with respect to the different RQs. Each RQ is associated to one of the main reference concepts of this paper, namely drivers, barriers and impacts. For each RQ, we report:
\begin{itemize}
    \item a \textbf{summary table} with categories associated to the concept, themes within a category, and codes within a theme;
    \item a list of the main \textbf{categories} (e.g., social, technical, etc.) identified for the specific topic;
    \item a textual \textit{explanation} of the themes within a category;
    \item a set of \textit{fragments} that exemplify the themes, tagged with the specific code (in square brackets).
\end{itemize}

\subsection{RQ1: What are the barriers hindering digitalisation in rural areas?}
\label{sec:barriers}

Barriers are reported in Table~\ref{tab:barriers} and are categorised into socio-cultural, technical, economic, environmental, and regulatory-institutional. Below, we discuss the different categories and internal themes. We report fragments of the interviews together with the codes associated to them, to provide evidence of the relation between data and themes. 

\begin{sidewaystable}
\begin{scriptsize}
    \centering
    \begin{tabular}{l|l}
    \hline
    \multicolumn{2}{c}{\textbf{Socio-Cultural Barriers}}                                      \\ \hline \hline
    \textbf{demographic} & age issues, social isolation, sparse population, seasonal work                                \\ \hline
    \textbf{distrust}    & distrust of funders, distrust of regulators, distrust of ICT supplier, distrust of technology \\ \hline
    \textbf{fear}        & fear of dependency from technology, fear of hidden costs, privacy concerns                    \\ \hline
    \textbf{values}      & attachment to tradition                                                                       \\ \hline
    \textbf{competence}  & lack of education, lack of knowledge, lack of skills, digital debt                            \\ \hline
    \textbf{complexity}  & complexity of regulations, complexity of technology, paradox of choice                        \\ \hline \hline
    
    \multicolumn{2}{c}{\textbf{Technical Barriers}}                               \\ \hline \hline
    \textbf{connectivity}  & absence of infrastructure, low quality of infrastructure  \\ \hline
    \textbf{dependability} & poor reliability, low efficiency \\ \hline
    \textbf{usability}     & poor ergonomic standards, poor usability in the field            \\ \hline
    \textbf{scalability}   & limited data storage,  limited computing capacity                      \\ \hline \hline
    
    \multicolumn{2}{c}{\textbf{Economic Barriers}}                                                                                    \\ \hline \hline
    \textbf{costs} & cost of technology, modernization cost, maintenance cost, lack of evidence of cost-effectiveness, lack of funding \\ \hline
    \textbf{scale} & small market size, small business size, atomized business structure  \\ \hline
    
    \multicolumn{2}{c}{\textbf{Regulatory-Institutional Barriers}}                                                                                    \\ \hline \hline
    \textbf{data management}          & {unclear data ownership, unclear data governance}                                                     \\ \hline
    \textbf{regulations}              & {frequent change of regulations, legal restrictions on technology, inadequate grant schemes criteria} \\ \hline 
    
    \end{tabular}
    \caption{Barriers hindering digitalisation in rural areas.}
    \label{tab:barriers}
\end{scriptsize}
\end{sidewaystable}

\paragraph{Socio-Cultural Barriers}
Most of the barriers to digitalisation are rooted in the cultural, socio-demographic, and somewhat emotional aspects and inclinations  of the individuals populating the rural communities. We identify six types of barriers: 
\begin{enumerate}
    \item \textit{demographic}, related to age issues, the logistic isolation of rural communities, the sparse, low-density population, and the presence of seasonal work, which makes rural areas places in which there is a limited permanent human presence for large part of the year. \smallskip \\ 
    \textbf{[demographic\footnote{This fragment is associated to all the codes in the demographic theme.}]}
\textit{Main limitations of these sectors are the atomized structure, the harsh working conditions, the seasonal work and the sparse and aged rural population. All of them facilitate the social and economic isolation, favoring the physical barriers.}
    \item \textit{distrust}, which is oriented towards different players, from founders and regulators, to ICT suppliers and technology in general. \smallskip \\ 
    \textbf{\textbf{[distrust of supplier]}} \textit{[There is] lack of trust in partners who use the data, which can be ICT companies (who may use the data for profiling, or on the stock market or who may sell the data) or other partners in the value chain (for example, if the farmers and the slaughterhouse start to share data, who will then harvest the benefits: the farmer or the slaughterhouse?)}

    \item \textit{fear}, often based or real threats, such as the risk of dependency from technology, the presence of hidden costs such as those related to maintenance of installed technology, and the privacy concerns related to data sharing. 
    \smallskip \\
    \textbf{[fear of dependency from technology]}
    \textit{Finally, there is also a fear of dependency and loss of control among some farmers. For example, by investing in monitoring systems farmers are increasing their dependence from management systems that need internet and electricity to successfully operate. Thus, sudden shocks like electricity loss might have devastating impacts on the farm.}
    \smallskip \\
    \textbf{[fear of hidden costs]}
    \textit{It might also be {no comma} that farmer decides not to implement the solution because of the challenges associated with the maintenance of the novelty.}
    
    \item \textit{values}, and in particular the attachment to traditional ways of working and identity. 
    \smallskip \\
    \textbf{[attachment to tradition]} \textit{Technology will require a higher level of digital education/training of the farmers. So far, many of them are reluctant to use a lot of technology as it does not fit to their image of being a farmer (e.g. working with the soil).} 
    
    \item \textit{competence}, such as general lack of higher education, specific knowledge of technologies, as well as practical skills to deal with technology, and, when these aspects become endemic, the emergence of \textit{digital debt} that increases the competence barrier to be covered. 
    \smallskip \\
    \textbf{[lack of knowledge]}
\textit{Another key challenge for farmers is to find staff that would have agricultural education yet would also have the knowledge regarding the cutting-edge farming software and hardware.}
\smallskip \\
\textbf{[digital debt]} 
\textit{Because of the poor material connectivity, people managed to cope without digital connectivity, and now they lack the ``digital capital'' to join the bigger leap in digitalization (using big data for business, using apps in their daily life, maintaining digital business  relations and so on). }
\smallskip \\    
    \item \textit{complexity}, which deals with the relationship between the individual and the feeling of being overwhelmed by the complex systems of regulations, the complexity of technology, and the paradox of choice due to the wide variety of technological solutions available in the market.
\smallskip \\    
\textbf{[paradox of choice]} 

\textit{As barriers: cost, complexity, skills and the fact that people are lost in the profusion of existing solutions. When farmers are talking about this to their advisors, the latter are sometimes as lost as farmers and limit themselves to propose solution they  control.}
\end{enumerate}

\paragraph{Technical Barriers}
Technical barriers are extremely relevant when it comes to digitalisation, and are related to four main quality aspects: 

\begin{enumerate}
    \item \textit{connectivity}, as the absence of a communication infrastructure in rural areas is one of the issues mentioned most often by the interviewed experts;
    \smallskip \\
    \textbf{[connectivity]} \textit{In my research I have seen that rural communities have been, and still are, on the wrong side of a digital divide. Over the past two decades this was mainly a material matter, with a lack of connectivity as the prime issue.}
    \smallskip \\
    \textbf{[connectivity]} \textit{The use of mobile applications allows us to reach a good part of the population and users of the territory, but unfortunately we cannot implement new tools without having the possibility of providing telephone and internet coverage to the entire territory.}
    \item \textit{dependability}, since, when present, technologies need to be dependable especially in particular environmental conditions such as those of fields and forests;
    \smallskip \\
    \textbf{[malfunctioning]} \textit{The agricultural environment is a relatively challenging environment. I have had that with the Near Infrared (NIR) sensors for manure tankers, I studied that, also in collaboration with companies. And that is a challenging environment, especially manure is very corrosive. So you really have to think about how you want to ensure the quality of your sensor technology over a longer period. Especially when driving in the field, with a lot of ammonia, yes, robustness is very important. }
    
    \item \textit{usability}, as standards required by the usage of a mobile phone in a field are not the same as those of the same device for daily usage; 
    \smallskip \\
    \textbf{[usability in the field]} \textit{[One of the promising technologies is the] use of natural languages recognition to facilitate the interactions with machine (e.g. manage crop operations and field log using voice interaction instead of manual entry).}
    
    \item \textit{scalability}, in terms of size and time complexity, since the amount of environmental data coming from monitoring systems is large, and need to be efficiently processed to take informed decisions in acceptable time, possibly profiting from edge computing solutions.
    \smallskip \\
    \textbf{[scalability]} \textit{until what point will we incorporate calculation treatment inside sensors? Today our computer models are based on ``cloud'' which means that farmers are locally collecting information thanks to sensors, smartphones or computers. Then raw data are sent to a distant server which will treat them, make calculations, cartographies and recommendations. After that, those results are sent back to farmers’ terminal. But cloud needs that raw information leave from the place they are so it needs a big communication effort between server and data collection area.}
\end{enumerate}

\paragraph{Economic Barriers}
Economic barriers are mostly related to the difficulty in dedicating financially sustainable investments in technological solutions, when margins are already limited as it happens in the primary sector. Main themes are:  
\begin{enumerate}
    \item \textit{costs}, including cost of technology but also cost of  modernization of the physical infrastructure of farms, low evidence of cost-effectiveness and the general lack of funds needed to afford the modernization. 
    \smallskip \\
    \textbf{[lack of funding, cost of technology]} \textit{By far the most significant barrier is funding. The technologies are expensive and not all farmers have the funds needed to cover the expenses.}
    \smallskip \\
    \textbf{[modernization cost]} \textit{Another important barrier is related to the properties of infrastructure. Installation of hardware needed to gather data for management systems or to install milking robots requires that farms correspond to certain characteristics. This might mean that farm building is too small, the ceiling is too low, the farm doors are too narrow, or some other solutions should be introduced before a farmer can incorporate the one he/she is aiming for. In these cases, the modernization is just too expensive and might include complete reconstruction of the farm.}
    \item \textit{scale}, as rural communities in EU are normally small business and do not have the mass to invest in costly technological renewals.  
    \smallskip \\
    \textbf{[small business size]}
    \textit{Furthermore, margins are often rather small/thin in rural businesses (small and micro family businesses often dominate the business landscape in rural areas) and this means that businesses can be caught up in trying to make break-even.}
    \smallskip \\
    \textbf{[small market size]}
    \textit{High value enterprises such as milk production will justify the technology many years before low value sectors such as lamb production.}      
\end{enumerate}

\paragraph{Regulatory-Institutional Barriers}
Institutions are also responsible for some barriers, as inadequate or unclear policies can hamper access to funds and technology. In particular, in relation to:

\begin{enumerate}
    \item \textit{data management}, which is often unclear in terms of who owns the data coming, e.g., from farm  monitoring systems and how these are managed.
    \smallskip \\
    \textbf{[unclear data ownership]} \textit{``Data food consortium'' [...] is about to develop a digital standard in order that all data can be integrated from one digital catalogue of products to another to decrease organizational costs and reinforce the control of data ownership. Farmers should only be able to give their agreement on data sharing for a precise and known use.}
    \item \textit{regulations}, which are frequently changing, and are sometimes not appropriate for rural contexts when it comes to grant schemes, which tend to privilege endeavours from large-size players. 
    \smallskip \\
    \textbf{[inadequate grant schemes criteria]} \textit{The EU funds is an important mean to overcome the challenges associated with access to funds. However, not everybody corresponds the criteria set by the grant schemes.}
\smallskip \\
\textbf{[inadequate grant schemes criteria]} \textit{My impression is that rural businesses are more regularly denied access to funding in comparison to urban counterparts. The adage seems to be ``scale up or quit''.}
\smallskip \\
\textbf{[frequent change of regulations]} 
\textit{The legislative context is ultra-changing so the one who says he want to revolutionize the word of agriculture and food industry in general will not succeed.}
\end{enumerate}

\subsection{RQ2: What are the drivers facilitating digitalisation in rural areas?}
\label{sec:drivers}

Drivers identified from the interview analysis are reported in Table~\ref{tab:drivers}, and are grouped into the same categories of barriers. The reader will already notice that while for drivers we have most of the themes in the economic and regulatory-institutional categories,  barriers are mostly socio-cultural and technical. Below, we report excerpts from the interviews, to highlight relevant drivers in each category.

\begin{sidewaystable}
\begin{scriptsize}
    \centering
    \begin{tabular}{l|p{0.75\textwidth}}
    \hline
    \multicolumn{2}{c}{\textbf{Socio-Cultural Drivers}}                                                                                                     \\ \hline
    \textbf{practical demands}   & demand for work flexibility, demand for workload reduction, demand for   wealth, demand for employment, need to reduce isolation \\ \hline
    \textbf{cultural tendencies} & cooperative spirit, solidarity spirit, need for inclusion,   technological fascination, trust in technology                   \\ \hline \hline
    \multicolumn{2}{c}{\textbf{Technical Drivers}}                                                                                                          \\ \hline \hline
    \textbf{quality}             & simplicity of technology, customisation of solutions, proven reliability, proven efficiency                                \\ \hline
    \textbf{service}             & more connectivity, availability of technology, data availability                                                           \\ \hline \hline
    \hline
    \multicolumn{2}{c}{\textbf{Economic Drivers}}                                                                                                                                    \\ \hline \hline
    \textbf{market demands}          & competition, consumer health concerns, green company image,   transparent company image, demand for certification, demand of organic   products \\ \hline
    \textbf{organisational}          & presence of intermediary roles, collective forms of organisation,   opportunity for cooperation                                                          \\ \hline
    \textbf{business needs}          & need for better control, need for simplification of legal   compliance, need for process optimization, need for better planning                 \\ \hline
    \textbf{financial}               & decreasing cost of technology, need for cost-effectiveness                                                                                      \\ \hline
    \textbf{labour}                  & shortage of labour, cost of manual labour                                                                                                       \\ \hline \hline
    \multicolumn{2}{c}{\textbf{Environmental Drivers}}                                                                                                                               \\ \hline \hline
    \textbf{impact reduction}  & need to reduce environmental impacts, need to reduce fertilizers, need to reduce pesticides                                                    \\ \hline
    \textbf{control}                 & need to decrease food waste, need to improve animal welfare, need to control natural disasters                                                  \\ \hline
    \multicolumn{2}{c}{\textbf{Regulatory-Institutional Drivers}}                                                                                                                    \\ \hline
    \textbf{regulatory restrictions} & taxes, constraints, need for regulatory compliance                                                                                                     \\ \hline
    \textbf{economic incentives}     & funding programmes, subsidies, incentives for technological   adoption, support for cooperation                                                 \\ \hline
    \textbf{educational support}             & training programmes, technical mentorship, support of education,   digital innovation centers                                                   \\ \hline
    \textbf{promotional}             & dissemination of results, promotion of digital entrepreneurship,   promotion of digital innovation                                              \\ \hline
    
    \end{tabular}
    \caption{Drivers facilitating digitalisation in rural areas.}
    \label{tab:drivers}

\end{scriptsize}
\end{sidewaystable}

\paragraph{Socio-cultural Drivers} 
Socio-cultural drivers include all those aspects that are related to the main social needs of rural communities  and to the typical inclinations and tendencies of stakeholders. We thus identify the following themes:

\begin{enumerate}
    \item \textit{practical demands} represent the social needs, and are related to: (i) reduction of isolation through better communication that can allow to identify and strengthen the links between needs and potential supply; (ii) demand for lighter work, as automation is expected to reduce the effort of manual labour typical of rural activities. 
    \smallskip \\
    \textbf{[need to reduce isolation]} \textit{New technologies break the existing isolation in those areas, providing the necessary communication coverage. [...] So the technology allows you to mix and match, it allows you to identify where the needs are and where there is a potential supply and to improve the links between them.}
    \smallskip \\
    \textbf{[demand for workload reduction]} \textit{And from a social point of view [...] many technologies reduce the workload for everyone, [such as automatic] steering systems and these are the drivers.}
    
    \item \textit{cultural tendencies}, which include the natural cooperative and solidarity spirit of small communities, the need for inclusion in the ``local vibe'', but also the fascination that technology can create in some of its users.
    \smallskip \\
\textbf{[solidarity spirit]} \textit{Historically [...] there is solidarity among neighbours and people who live in rural areas, and what the digital does is to allow that natural resilience and solidarity to come out more and it facilitates it. }
\smallskip \\
\textbf{[need for inclusion]} \textit{With community members [...] the driver seems to be to get included, to join others in (online) groups, and make sure one stays part of the local vibe. One of the issues in running a business, also in rural areas, is to make sure to also include those who do not feel the urge to go `digital'. This means that drivers for digital adoption go beyond mere monetary cost-effectiveness. Social drivers such as inclusion, but also comparing with peers (other businesses) often turn out to be straightforward, and old-fashioned if you like, motivational factors.} 
\smallskip \\
\textbf{[technological fascination]} \textit{When it comes to harvesting for big crops like barley, wheat etc there are these large scale harvesting machines that workers use for many days in a row. After interviewing farmers in Denmark, that have used yield monitoring tools on their machinery, they stated that their daily job has become more interesting. So we can assume that there is also a social aspect that comes with the use of digital tools in agriculture.}
    
\end{enumerate}

% Part of the drivers derive from \textit{practical demands} of the rural community, including business owners, workers and citizens. 

\paragraph{Technical Drivers}
A limited yet relevant part of the drivers is also technical, intended as relevant non-functional attributes of technology that can play a crucial role in facilitating digitalisation. We identify two main themes:

\begin{enumerate}
    \item \textit{quality}, intended as evidence about the proven quality of the technology in terms of simplicity, reliability, efficiency, and the possibility to adapt it to the peculiarities of rural areas can convince rural community members to accept digital solutions.
    \smallskip \\
\textbf{[simplicity of technology]} \textit{In general, the development is much more driven by technology, economics, and acceptance by farmers than by policy. Acceptance is strongly driven by the simplicity or complexity in using the technology. E.g. simple smart phone applications are much easier accepted and used than complex systems requiring PC and specific training.}
    \smallskip \\
\textbf{[customisation of solutions]} \textit{The uptake of the technology is also facilitated by the fact that there are new versions of technologies constantly being developed aimed at specific subgroups of farmers (e.g. small dairy farms).} 

    \item \textit{service}, related to the availability of certain ICT service---e.g., basic connectivity or technology---, which can facilitate digitalisation just thanks to the mere possibility of being accessible by rural area stakeholders. Similarly, digitalisation can be fostered by the availability of new types of data about plants and crops that can be exploited for better monitoring and control. 
     \smallskip \\
\textbf{[more connectivity]} \textit{First of all, you do need the technology. So the connectivity has to be there.}
 \smallskip \\
\textbf{[data availability]} \textit{The last area that has emerged derived from the vast amount of data and images that we have collected from different crops, where we are trying with deep learning techniques to explain some of the features of this collected data.}
    
\end{enumerate}

\paragraph{Economic Drivers}
The largest part of drivers is economic and business-related, and we identify  five main themes in this category: 
\begin{enumerate}
    \item \textit{market demands}, collecting drivers coming from customers' requests for healthy food and market trends, such as the need to have a ``green'' and transparent image. The demand, e.g., for organic products is related to health concerns, but becomes also a powerful market opportunity as ``green'' becomes fashionable.   
\smallskip \\
\textbf{[consumer health concerns]} \textit{Whereas before we gave more fertilizer than required, but now we see that this is no longer possible. And now the margin of fertilizer is reduced, and to prevent yield losses we need to efficiently apply fertilizer and drive the need for innovation. And similarly, we see a need to reduce the use of pesticides. The consumer wants less inputs while we do need to be able to protect the crops, and we see a driving force for innovation there, driven by society.} 

\item \textit{organisational}, with drivers related to novel organisational structures. These includes collective forms of organisation such as cooperatives that can facilitate small players, but also technological hubs and intermediary roles, facilitating learning and access to technology. 
\smallskip \\
\textbf{[presence of intermediary roles]} \textit{I think there are critical intermediaries that play a vital role. They take different forms in different countries. So, you have different types of digital hubs, you have co-working spaces, you have Fab Labs, you have virtual labs of different kinds. These intermediaries play a vital interface role between the people who are using the technology and the providers of it. }
\item \textit{business needs}---internal company needs, such as process optimization; 
\smallskip \\
\textbf{[need for better planning]}
\textit{The adaptation of these technologies mainly depends on the economic possibilities of a particular farm. They are introduced to improve the efficiency of the farm – in terms of higher cow productivity, more efficient reproduction planning, reduced calf mortality and other calf related challenges, etc.}
\item \textit{financial}, collecting phenomena related to cost of assets and expected benefits. As advanced technology becomes less expensive, more subjects can take the risk of experimenting with technologies. 
\smallskip \\
\textbf{[need for cost-effectiveness]}
\textit{In this line, the use of Unmanned Aerial Vehicle (UAV) on-board solutions for the analysis of agro-environmental phenomena is also being implemented very quickly, due to the possibilities of use at a detailed scale and at an affordable cost compared to other traditional techniques.} 
\item \textit{labour-related}, including phenomena related to shortage or cost of labour, as, on one hand, rural areas tend to be scarcely populated, and on the other hand the cost of manual labour can be too high with respect to the typical revenues of the primary sector. \smallskip \\
\textbf{[shortage of labour]} \textit{Milking robots are being installed to counter the labour shortages and to improve the efficiency of farms.}
\end{enumerate}

\paragraph{Environmental Drivers}
Sustainability is strictly related with the needs of a silent stakeholder, namely the environment. Some drivers are therefore concerned with the relationship between the subject and the ecosystem. These are grouped into two somewhat mirror categories: 

\begin{enumerate}
    \item \textit{impact reduction}, collecting drivers related to the need to  reduce human impacts, in terms of reduced usage of fertilizers that can harm the soil in the long terms, and in terms of less pesticides, which disrupt biodiversity. 
    \smallskip \\
\textbf{[need to reduce fertilizers]} \textit{Whereas before we gave more fertilizer than required, but now we see that this is no longer possible. And now the margin of fertilizer is reduced, and to prevent yield losses we need to efficiently apply fertilizer and drive the need for innovation.} 
    \item \textit{control}, which are drivers concerned with the need to control the environment, such as the need to improve animal health, and control natural disasters.
    \smallskip \\
\textbf{[need to improve animal welfare]} \textit{AMS may have significant potential in the prevention of adverse health outcomes in milking of dairy cows in comparison to conventional milking systems.}
\smallskip \\
\textbf{[control of natural disasters]}
\textit{All the tools currently used [...] for the analysis and monitoring of environmental phenomena, such as satellite images, orthophotos, LiDAR data, UAV, are already being used in agriculture, forestry and rural areas. [...] They are very plastic solutions adapted to the control and monitoring of key parameters of different production systems and to the prevention and control of natural disasters.}
\end{enumerate}

\paragraph{Regulatory-Institutional Drivers}
Institutions are the actors that can contribute the most to the digital transformation, using different policy instruments that can steer the direction of the rural  communities. The main instruments are: 

\begin{enumerate}
    \item \textit{regulatory restrictions}, such as new regulations, with taxes and constraints associated to undesired behaviours, as, for example, the excessive usage of nitrogen for fertilization. 
\smallskip \\
\textbf{[need for regulatory compliance]}
\textit{Required reduction of the nitrogen balance in the new agricultural policy (AP 2020) will certainly increase the interest in carrying out nitrogen fertilization more precisely and using the available nitrogen as optimally as possible.}

\item \textit{economic incentives}, such as funding programmes, subsidies, incentives for the adoption sustainable technologies, and economic support for cooperation with digital players. 
\smallskip \\ 
\textbf{[funding programmes]} \textit{In the forestry sector, the RDP funding is the main factor acting as a driver from both an economic and a social point of view.}
\smallskip \\ 
\textbf{[incentives for technological adoption]} 
\textit{Finally, another important aspect is having policies that will incentivize people to adopt new technologies. These policies range from, public awareness, taxes and subsidies, training and education\footnote{This fragment has been coded also under the theme \textit{educational}, with the code ``training programmes''.}, cohesion funds and in general policies that aim to shift the risk away from the technology user can become a driving force in ICT adoption.}

\item \textit{educational support}, to facilitate the circulation of digital knowledge with training programmes, technical mentorship and the creation of digital innovation centers.
\smallskip \\ 
\textbf{[digital innovation centres]}
\textit{The creation and development of digital innovation centres, specifically in the agri-food sector (Agri Food DIH) [are relevant drivers]}

\item \textit{promotional}, with campaigns oriented to promote digital innovation and disseminate results of success stories.
\smallskip \\ 
\textbf{[promotion of digital entrepreneurship]}
\textit{the promotion of digital entrepreneurship through conferences and seminars and demonstration activities}
\end{enumerate}

\subsection{RQ3: What is the potential impact of digitalisation in rural areas?}

When asked about the potential impacts of digitalisation, the experts discussed cases of positive and negative ones, both based on their previous real-world experience and on speculation. Table~\ref{tab:impact} summarises the results. We identify four main categories. Most of the identified themes and codes have a strong social angle, and we considered it appropriate to reflect this in the category names:  socio-cultural, socio-economic, socio-political and environmental. 

Each theme reported in Table~\ref{tab:impact} is also linked to the main stakeholders affected by a certain type of impacts. We identify five, non-exclusive, classes of stakeholders, namely: 
\begin{itemize}
    \item the \textbf{community} as a whole social subject; the workers, employed in farms and in other businesses; 
    \item the \textbf{business owners}, distinguished into small and large, depending on the size of the business, as this has an effect on the type of impacts; 
    \item the \textbf{institutions}, intended as municipalities, but also region, states, regulators and policy-makers in general; 
    \item the \textbf{environment}, which is again a key stakeholder that sustainable development needs to take into account.  
\end{itemize}

As the analysis of impacts is multi-dimensional, as it accounts for positive and negative impacts, as well as different stakeholders. Therefore, to highlight relevant relationships across dimensions, we discuss our findings for this RQ is an argumentative manner, instead of linearly summarizing each single theme, representative fragments at the bottom of the description of each category.

\begin{sidewaystable}
    \begin{scriptsize}

    \centering
    \begin{tabular}{p{0.2\textwidth}|p{0.35\textwidth}|p{0.35\textwidth}} 
    \hline & \textbf{\textit{Positive}} & \textbf{\textit{Negative}} \\ \hline \hline
    
    \multicolumn{3}{c}{\textbf{Socio-Cultural Impacts}} \\ \hline
    \textbf{social} (community) & inclusion, reduction of disadvantages, involvement of consumers, improved attractiveness of rural areas &  exclusion \\ \hline
    \textbf{quality of life} (community) & more free time, increased access to external goods, improved well-being, relief from heavy work & detachment from nature \\ \hline
    \textbf{education} (community) & access to distant learning, increased education & closing of local schools, loss of expertise \\ \hline  \hline
   
    \multicolumn{3}{c}{\textbf{Socio-Economic Impacts}} \\ \hline  \hline
    \textbf{labour} (workers, business owners) & replacement of repetitive labour, replacement of seasonal labour, shift to technology-based labour, novel job opportunities, better access to skilled workforce, decentralization of work structure & unemployment, change in work profiles \\ \hline
    {\textbf{financial} (business owners)} & {more profits, reduction of costs, improved   productivity, benefits of scale} & - \\ \hline
    {\textbf{management} (\textit{large} business owners)} & {control at larger scale, optimization of resources, increase of business choices, better management of process, logistic optimized, better management of production irregularity, improved measurability} & change of stakeholders, change in production models \\ \hline
    {\textbf{market} (community, \textit{small} business owners)} & {improved tourism, novel energy-related services, attraction for technology players} & loss of independent companies, increased performance inequality, closing of  local businesses, creation of monopolies, interest   of large subjects, scale effects on small farms,  increased dependency on global markets \\ \hline  \hline
    
    \multicolumn{3}{c}{\textbf{Socio-Political Impacts}} \\ \hline  \hline
    {\textbf{data} (business owners)} & {power control in data management, increased control of data ownership, improved transparency, improved trust, increased value of data} & - \\ \hline
    {\textbf{institutional} (community, institutions)} & {improved food democracy, improved legality, facilitated regulatory compliance} & - \\ \hline \hline
    
    \multicolumn{3}{c}{\textbf{Environmental Impacts}} \\ \hline \hline
    \textbf{environment} (environment) & reduction of human impacts, improved sustainability, reduction of carbon emission, improved animal welfare & - \\ \hline
    \end{tabular}
    \caption{Expected impacts of digitalisation in rural areas.}
    \label{tab:impact}
    \end{scriptsize}
\end{sidewaystable}

\paragraph{Socio-Cultural Impacts}
Expected impacts on the community as a whole is concerned with three main themes: \textit{social} and relational aspects, with higher inclusion of rural areas into the society at large, and improved attractiveness of rural areas; the general \textit{quality of life}, due to the relief from heavy work that can give access to more free time, but also to the possibility of accessing goods from distant areas through online purchases; \textit{education}, with the availability of distant learning and increased education driven by the need to learn the technology itself.

Digitalisation comes also with its risks, such as the exclusion of those subjects who cannot keep the pace of technological change, but also the detachment from nature, since the relationship between workers, fields and animals is increasingly mediated by computers and robot. Furthermore, access to distant learning can lead to the closing of local schools, while the increase in digital automation can lead farmers to lose their expertise and intuition, as they would rely more and more on data analysis and decision making systems. Fragments exemplifying these positive and negative impacts are reported in the following. 

\textbf{[social: inclusion]}
\textit{And I think this [technological intermediaries] helps people to step up in a progressive way. So there's the trends, the digital journey, if you like. And then the idea that rural areas are not alone. They're part of something bigger, and they need to work out how they link in with them.}

\textbf{[social: exclusion]} 
\textit{There is a real need of farmers' education about the use of digital and increased intelligence. But tools must also be reliable, ergonomic, trusted and couple together with a training for their use. A farmer who will be excluded from digital will be excluded from the system.}

\textbf{[quality of life: more free time]} \textit{Finally, these solutions simplify the daily life of farmers and, presumably increases their quality of life. Many of these farmers did not have time for anything before they modernized the farm. Modernization was a way to ensure that there is time for off-farm activities. } 

\textbf{[quality of life: detachment from nature]} \textit{[A risk is] less human attention to animals and plants. Too much to rely on programming and machinery---not all situations can be predicted and programmed.} 

\textbf{[education: access to distant learning, closing of local schools]} 
\textit{it is clear that distant learning can be of help to students in villages and in more rural areas, but at the same time it can provide an excuse to closing down the village school and concentrate the village schools in other places}

\textbf{[education: loss of expertise]} 
\textit{it could lead to a loss of expertise and common knowledge of farmers if robots are more used to determine production decisions.}

\paragraph{Socio-Economic Impacts}
The largest part of the discussed themes are related to the socio-economic impacts of digitalization, and in particular concerning four main aspects that affect workers, business owners and the community. Impacts in this category are mostly positive for what concerns \textit{labour}, \textit{financial} aspects and \textit{management} aspects, while more disruptive changes affect the \textit{market} models, especially when considering effects on small players. 

Concerning \textit{labour}, positive impact is the replacement of repetitive and seasonal labour, the presence of novel job opportunities associated with the usage of new technologies, but also the possibility of exploiting the network to gain access to a skilled workforce and decentralise the work structure. Undesired impact is mainly the possibility of unemployment, but also the need to cope with the change in work profiles. 

\textbf{[labour: replacement of repetitive labour]} 
\textit{And especially you will see a shift between low-skilled, repetitive labor to more technology-based labor where the dirty work is done by the machine. And the oversight and interpretation is still a task for people. That also means that the education becomes more and more important.}

\textbf{[labour: unemployment]}  \textit{We can think that robots will be able to alleviate heavy work and to overcome the difficulty to find working force when they will be more accessible. But on the other hand, some employees will not have work anymore.}

Financial aspects are generally positive for business owners, with more profits, reduction of costs, improved productivity especially thanks to the usage of data acquisition and monitoring systems, and the possibility of leveraging technology to scale-up with the same amount of labour. 

\textbf{[financial: improved productivity]} \textit{About the question of performances, a study was made few years ago on the use of milking robots. This work shows how diverse was the use of a same tool among farmers, ranging from simple milking to enhance generated data during milking for operations management. It also shows that a great productivity gain was made with the intensive use of the digital data and a pretty modest gain for those who used it as a simple milking tool.}

From the \textit{management} standpoint, the main beneficiaries are the large business owners, who can achieve better control at larger scale, optimize their resources and processes, deal with production irregularity thanks to the improved measurability granted by the sensing and monitoring technologies as well as the farm management platforms. 
Negative aspects are again concerned with the need to deal with change, in terms of production models, as the introduction of new tools require adjustments in the processes, and the change of involved stakeholders, with the strong presence of the technology providers.

\textbf{[management: control at larger scale]} 
\textit{Also the monitoring is automated with this technology. And that will mean, that when farms get bigger they can still keep an overview of the farm.} 

\textbf{[management: better management of production irregularities]} 
\textit{Digital can help to manage production irregularity. If the farmer has an alert all along the production’s process, he can adjust his position on the sector with more adapted specifications and better valorize his product on the final market.}

\textbf{[management: change of stakeholders, change of production models]} \textit{Digital technologies are not just ``tools'' added to a farm; they thoroughly change farm management and practice. They demand therefore a revision of the actions of farmers and the interaction with stakeholders around it. Moreover, it also changes the types of stakeholders who are part of the social network around farms. So, the innovation is not only a technological affair, but also a social one.}

Finally, the \textit{market} also sees positive changes thanks to the availability of online booking services and the birth of novel energy-related services, e.g., with the usage of renewable sources. However, the effect on the market can be particularly negative for small business owners, with the closing of local business who cannot compete on the global market, the increased performance inequality with respect to large players who can profit from technology, and the creation of monopolies, as the digital world is characterised by a tendency towards these types of centralised market models.  

\textbf{[market: closing of local businesses]} \textit{online shopping can help rural communities obtain lots of goods without travelling, that they would not have been able to do beforehand, but at the same time it can mean that a shop in the village loses customers and has to close down. }

\textbf{[market: creation of monopolies]} 
\textit{In addition, digital platforms as part of those ecosystems are usually having a disruptive impact as they follow a ``first player wins the whole market'' scenario, which is usually disruptive for other branches.}

\paragraph{Socio-Political Impacts}
The socio-political themes and codes identified are all related to positive effects, and have an impact on business owners, for what concerns aspects related to data, and their value and control, and to the community and institutions, with improved legality thanks to mechanisms that facilitate transparency and regulatory compliance, with the setup of blockchain systems for traceability, as well as dematerialisation and standardisation of processes.  

\textbf{[data: improved transparency]} 
\textit{Blockchain is not revolutionary but it secures data access. It’s a tool and all the issue is to give the feeling of protection, data security, monetarize it and give access.}

\textbf{[institutional: improved legality]} \textit{The use of technology could also foster greater transparency at global level, due to the adoption of digital tools in Countries that are net exporters of fuelwood (such as Bosnia - Herzegovina and Ukraine). In case of adoption of these new technologies (i.e., blockchain), as a consequence companies from these Countries could enter new (legal) markets thanks to their compliance with the European Union Timber Regulation (EUTR). In turn, Italian companies that import wood would benefit from a large simplification of their Due Diligence System (DDS).}

\paragraph{Environmental Impacts} The environment is assumed to benefit from the introduction of digitalization and technological solutions, with reduction of human impacts, carbon emissions and improved animal welfare, while no risks were explicitly mentioned by the experts. The expected environmental impacts are strictly linked to the environmental drivers, which were previously discussed, and therefore we report only one representative fragment in this theme. 

\textbf{[reduction of carbon emission]} \textit{Digital allows to bring data together and sharing them optimize logistic and limit carbon  emissions, cost and mobility.}
\section{Discussion}
\label{sec:discussion}

The analysis of the interviews specialises the conceptual framework outlined in Sect.~\ref{sec:framework} for the domain of rural areas, based on information elicited from domain-experts from the ICT and social-science sector. RE practitioners involved in the development of technological applications in this domain shall consider the list of themes identified, and incorporate them in their RE activities, e.g., in the form of checklists to make evident which sustainability concerns are considered in the requirements specifications, or by mapping them into interview scripts for sustainability requirements elicitation. A specific methodology with detailed guidelines is out of the scope of the current paper, and is left as future research. In the following, instead, we summarise identified barriers, drivers and impacts, and we relate them with previous literature. Our findings shall be considered by RE practitioners when dealing with system development in rural areas, such as applications in digital farming~\citep{bacco2019digitisation}, smart villages~\citep{doerr2018reinrural}, smart forestry~\citep{zou2019survey} and similar.

%highlight what are the main implications of our findings for RE research and practice.  

\paragraph{Barriers} Barriers to digitalisation in rural areas are mostly \textit{socio-cultural}, and in particular demographic issues related to aging and sparse population. Technology requires skilled users that are less frequent among older people, as well as exchange of technological knowledge, which is made difficult in a highly distributed and scarcely populated context. Demographic factors were already observed to be crucial aspects for technology adoption by the literature review on precision agriculture by~\cite{pierpaoli2013drivers}, as well as the more recent survey by Paustian~\cite{paustian2017adoption}, and our findings confirm this vision. 

Rural communities also tend to rely on traditional values, and having negative sentiments towards novelty. These sentiments include  \textit{distrust}--- especially towards all those parties that are regarded as external to the rural environment, namely funders, regulators, and ICT suppliers---and \textit{fear}. This is directed towards the  concrete possibility of becoming dependent from the technology, but also of unknown angles, such as hidden costs of technology and data ownership. Negative sentiments are not helped by the inherent complexity of technology and regulations, and by the lack of ICT skills in rural areas. The relevance of traditional values and their potential negative influence on technology adoption was confirmed by~\cite{regan2019smart} in an interview study with smart farming experts in Ireland. Educational barriers in terms of lack of training were also noted in the past, e.g., by~\cite{robertson2007opportunities} in a study on precision agriculture, and still appear to be relevant according to our experts. Finally, the need to address trust issues were remarked by~\cite{van2019ethics}, in a study about ethics in smart farming. 

Technological barriers are also particularly relevant, and, most of all, \textit{connectivity}. Without internet connection, isolation is amplified as well as the possibility of using all those sensing technologies that rely on connectivity to properly function, such as cloud-based sensor networks and IoT platforms in general, as observed by~\cite{bacco2019digitisation} in a survey of digital technologies for smart farming. 
Other relevant technical barriers are to the need to adapt technologies to ergonomic requirements of  fields and forests. These require rugged devices that can be operated with gloves, possibly through voice interaction, and that resist to corrosion and adverse environmental conditions in general. While studies exist on ergonomic for agriculture that take into account physical tools and machines (see, e.g., the recent survey by~\cite{benos2020review}), we are not aware of similar works in RE for digital systems. 

Economic barriers are the \textit{costs} of technology adoption, as well as scarce evidence of cost-effectiveness of certain solutions. Rural areas are often characterised by small players and atomized business structures, which cannot take advantage of the economies of scale favoured by the deployment of a  technological infrastructure. The challenge of cost was also highlighted by~\cite{bacco2019digitisation} and by~\cite{barnes2019exploring}, who identify it as first barrier for the adoption of precision agriculture. Instead, the need to provide evidence of the return of investment in specific contexts was noted by an earlier study of~\cite{barnes2018influencing} and~\cite{regan2019smart}.

Finally, barriers are also \textit{regulatory-institutional}, especially in relation to data management. Issues around data ownership were widely discussed also by previous literature focused on smart farming and the agrifood sector in general~\citep{barnes2018influencing,regan2019smart,van2019ethics,fleming2018big,schroeder2021s,de2016ownership}. As observed by~\cite{van2019ethics}, farm data are not personal data in strict sense, still they are valued so by farmers because farm business and household are traditionally viewed as `one-and-the-same economic unit'. The data ownership concerns of farmers, and their observed distrust towards regulators and ICT providers, are indeed exacerbated by the absence of clear policies for the management of farm data that can be collected by digital platforms. ~\cite{carbonell2016ethics} suggests to adopt open data policies, as a way to respect the people's right to access information. Still,~\cite{schroeder2021s} remarks the need for transparent data policies, in which the usage of data is made clear to farmers, and technology providers are accountable for how they use them. 

\begin{mdframed}[style=style1]
\faLightbulbO~~The main barriers for adoption of digital technologies are socio-cultural, and especially demographic and educational issues. Connectivity is the most important technical concern. Cost of technology adoption is the main economic barrier. Unclear data ownership is the main regulatory-institutional challenge. 
\end{mdframed}

\paragraph{Drivers} While the majority of barriers are socio-cultural, main drivers are \textit{economic}. The primary relevance of economic aspects was also observed in previous studies focused on precision agriculture~\citep{barnes2019exploring}, and it is evident also from the analysis of policy documents by~\cite{lajoie2020future}. Technology is expected to facilitate access to fine grained information about resources, e.g., soil, plants and animals, to address \textit{business needs} such as greater control of production, better optimization and better planning. On the other hand, technology is pushed also by external factors, such as \textit{market demands} for higher competition, but also for greener and transparent image. This can be facilitated by technologies for precision agriculture oriented to use less fertilisers, and other digital means, e.g., blockchain, to support food and wood traceability. Other important drivers are \textit{organisational}, as there are forms of organisation such as cooperatives that can facilitate small players in sharing the cost of technological change. In addition to that, technology mediators can facilitate access to novel digital solutions. The relevance of mediators, and in particular advisory services, was observed to be crucial by~\cite{busse2014innovation} in an interview study on precision farming in Germany.

Economic drivers are complemented by \textit{regulatory-institutional} ones, as restrictions on the usage of certain fertilisers, together with the need for product certification, pushes farmers to introduce technologies that provide evidence that regulations are respected.  
Economic incentives for technology adoption coming from institutions are also key enablers, paired with educational support by means, e.g., of digital innovation centres, which can facilitate technological uptake. The importance of subsidies and taxation as main enablers was also observed by~\cite{barnes2019exploring}, and the role of institutions in general is remarked also in a recent book by~\cite{schroeder2021s} focused on digitalisation in agrifood. 

Regulatory-institutional drivers are tightly connected with \textit{environmental} ones, with the need of reducing the impact of the human footprint, counterbalanced by the urge to better control the environment, for example from natural disasters. Overall, environmental drivers have been observed to be considered as secondary aspects, as primary concerns are related to profitability for farmers~\citep{barnes2018influencing}, and even for policy makers~\citep{lajoie2020future}. 

Some important factors are also \textit{socio-cultural}. As observed, some people resist to change, while others are technological enthusiasts, and can play the role of `technology sponsors', which can be effective in a community where the need for inclusion is high. The presence of these enthusiast is confirmed by the study of~\cite{paustian2017adoption}, who performed a questionnaire with German farmers, and showed that people with low experience in crop farming---less than 5 years, supposedly younger people---were more inclined to adopt smart farming technologies. Practical needs such as reducing heavy workload and improve work flexibility are also drivers for the technological change.  
\begin{mdframed}[style=style1]
\faLightbulbO~~Main drivers are economic, and the need to increase revenue and control of production. They are paired with regulatory-institutional drivers, including taxes, subsidies, economic incentives and the diffusion of digital innovation centers. Socio-cultural drivers come from young technology enthusiast.  Less explicit are environmental drivers. 
\end{mdframed}

\paragraph{Impacts and Entities} Discussed impacts have a strong social dimensions, with \textit{socio-economic} aspects dominating the scene. Business owners can improve their productivity by achieving better management of their processes, and by lowering the costs of labour through the replacement of manual activities with automatic ones. Large business owners can also take advantage of the economies of scale facilitated by technological infrastructures. On the other hand, while large players tend to be privileged by digitalisation, small ones risk to be ruled-out by a market in which they cannot compete, or to be incorporated by larger companies. Communities can suffer from these phenomena with the closing of local businesses, and with unemployment. The disparity between small and big players, accompanied by the risk of inequitable development is a topic of discussion in the literature~\citep{regan2019smart,fleming2018big}. The size of a farm was observed to positively influence the adoption of precision agriculture~\citep{pierpaoli2013drivers}, and the logistic regression analysis by~\cite{paustian2017adoption} empirically confirmed this intuition, showing that technology adopters tend to have more than 500 hectares of arable lands.

From the \textit{socio-cultural} perspective, positive expected impacts affect the rural communities as a whole. Among them are the increased inclusion, mostly driven by connectivity, and the greater well-being due to more free time and less heavy work. The mere presence of internet connection can make rural areas more attractive, and can facilitate access to distant learning to acquire the missing skills needed to introduce novel technologies. On the other hand, the risk of exclusion for those subject that cannot or do not want to use technology is high. In addition, a community that uses digital means as interface to the environment risk to lose expertise, and to get detached from nature. Positive socio-cultural impacts are confirmed by the survey of~\cite{regan2019smart}, especially in terms greater well-being achieved by reduction of burdensome jobs, and improved time management. The same study also confirms our findings that negative impacts include the over-reliance on technology, with consequent loss of skills, as well as the potential distancing and isolation of farmers from animals and community.

Other observed impacts are at the \textit{socio-political} level. These favour large business owners, the community and institutions alike. For example, improved control on data about a certain farm, or about the origin of wood, can facilitate assessment by government and thus improve legality, as remarked by other studies~\citep{schroeder2021s}. Finally, from the \textit{environmental} standpoint, technology facilitates precision and control, thus reducing the human impact on vegetation and animals in the long-term.

\begin{mdframed}[style=style1]
\faLightbulbO~~Main impacts are socio-economic, with replacement of manual work and the possibility to leverage economies of scale. Socio-cultural impacts include greater well-being, improved technical skills, but also loss of practical expertise due to dependency from technology.     
Positively impacted entities are large business owners, the natural environment, and institutions.  Negatively impacted ones are small players and manual workers who risk unemployment.
\end{mdframed}

\section{Conclusion}
\label{sec:conclusion}
Sustainable requirements are quality concerns that requirements engineers shall take into account when transforming existing socio-cyber-physical contexts through the introduction of novel digital technologies. Sustainability requirements involve mid- to long-term effects that the system can have on the context in which it is deployed. Previous research studied these requirements form the viewpoint of requirements engineering (RE) professionals~\citep{condori2018characterizing,chitchyan2016sustainability}. In this paper, we perform an interview study involving 30 sustainability experts with background in ICT and social-science in the domain of rural areas. Our study aims to elicit \textit{drivers}, \textit{barriers} and \textit{impacts} of digitalisation in rural areas, considered according to different sustainability dimensions. These concepts are regarded as having a wider temporal perspective with respect to stakeholders' goals normally discussed in RE, and we consider them to be the right lenses to analyse sustainability aspects. From the analysis of the interviews, we classify 14  barriers, 15 drivers, and 10 types of impact, divided into different sustainability categories adapted from previous literature~\citep{lago2015framing,duboc2020requirements,becker2015sustainability,goodland1995concept}, e.g., socio-cultural, economic, regulatory-institutional, environmental, \textit{etc.} According to the experts, the main barriers are socio-cultural, drivers are mostly economic, and impacts are balanced between social and economic aspects. Our findings can be useful to RE practitioners that deal with system development in rural areas, such as applications in digital farming~\citep{bacco2019digitisation}, smart villages~\citep{doerr2018reinrural}, smart forestry~\citep{zou2019survey} and similar. 

This paper is part of a larger endeavour carried out within the H2020 DESIRA project. In future work, we plan to confirm and extend the provided conceptual framework through the analysis of interviews and focus groups coming from the DESIRA Living Labs, which involve farmers and other rural area stakeholders. Particular attention will be devoted to the identification of \textit{negative impacts} and \textit{environmental} aspects in general. At this stage, we were not able to sufficiently foster discussion around these aspects, while we believe that a more concrete perspective, such as the one available in the Living Labs, can facilitate the enrichment of the framework in this direction.

\section*{Acknowledgement}
The following DESIRA partners contributed in carrying out and transcribing the experts' interviews: James Hutton Institute; FiBL; SISTEMA GmbH; Universidad de Cordoba; AEIDL; Zemnieku Saeima; Karlsruhe Institute of Technology ITAS; AMIGO srl; Wageningen Research; Wageningen University \& Research; Nodibinajums Baltic Studies Centre; Uniwersytet Lodzki; Flanders Research Institute for Agriculture, Fisheries and Food; Jyvaskylan Yliopisto; INRA; ATHENA; UNIDEB; PEFC Italy; Fraunhofer.

\bibliographystyle{unsrtnat}
\bibliography{bibliography}  %%% Uncomment this line and comment out the ``thebibliography'' section below to use the external .bib file (using bibtex) .





% \end{thebibliography}

\end{document}